\documentclass[twocolumn]{jpsj3}
\usepackage{txfonts,amsmath}
\newcommand{\dbar}{%
        \mbox{$\textrm{d}$\kern-0.28em\raise.7ex\hbox{-}}}

\title{A Numerical Formulation to Calculate the Conductance of Mesoscopic Conductors \\
Using Singular Value Decomposition}
\author{Masahiko Hayashi\thanks{E-mail: m-hayashi@ed.akita-u.ac.jp}}
\date{\today}
\inst{Faculty of Education and Human Studies, 
Akita University, Akita 010-8502, Japan}

\abst{
We present a new formulation to calculate the electric conductance of 
mesoscopic conductors by utilizing the singular value decomposition, 
which is a mathematical technique to manipulate matrices. 
Our formulation is useful in treating conductors with rather complicated 
atomic structures, for which naive recursion formula is cumbersome. 
It also has an advantage in scaling up the calculation by using parallel computation, 
which potentially allows us the real-scale calculations at the atomic level. 
On the other hands, the effects of electron-electron interactions are hard to be treated within this framework, 
since it depends crucially on the linearity of the system. 
In this paper, we study graphene nanoribbons with external leads for a simple example. 
}
\kword{Transport, Conductance, Landauer Formula, Graphene, Nano-constrictions}

\begin{document}

\maketitle

\section{Introduction}

Numerical calculation is one of the most powerful 
tools to study mesoscopic systems. 
Especially, the effects of impurities and randomness on 
the electric conduction have been intensively studied in the light of the Anderson localization 
\cite{Sarker:1980ug,Thouless:1981vz,Lee:1981ve,Pichard:1981wa,Baranger:1991ta}, 
and the recursive calculation method of the scattering states 
of electrons has been developed 
\cite{Thouless:1981vz,Lee:1981ve,MacKinnon:1985vc,Ando:1985vn}. 
These numerical approaches are strengthened by 
the Landauer formula\cite{Landauer:1957:SVC:1661287.1661290}, 
by which one can reduce the calculation of the conductance 
to the calculation of the scattering matrix of the conductor 
\cite{Fisher:1981tt,Buttiker:1985vk}. 
The so-called recursive Green function method 
has been established as a standard to calculate mesoscopic conductance 
\cite{Ando:1991wt}. 
This method has been extended to multi-terminal geometries 
in order to calculate the Hall conductance\cite{Baranger:1991ta}. 
Until now many sophisticated numerical algorithms have been developed 
to be applied to molecular or nano-devices 
\cite{Brandbyge:2002ck, Rocha:2006fk, Birner:2007gm, Ozaki:2010ec, 
Darancet:2010dy,Fonseca:2013bg, Groth:2014ia}. 

Recently, many people have strong interests in the transport properties of graphene 
\cite{Novoselov:2004it,Zhang:2005gp,Novoselov:2005es,Ando:2007eq}. 
The conductance through the graphene nano-wires has been studied, 
taking into account the effects of impurities, sample edges, and so on
\cite{Wakabayashi:1999ti,Wakabayashi:2001uv,Wakabayashi:2002jb,Peres:2006ju,Li:2008js,
Darancet:2009dw,Mucciolo:2009fy,Nakanishi:2010bb,Ihnatsenka:2012dh,
Kleftogiannis:2013cj,Ihnatsenka:2013cq,Takashima:2014du}.  
The experimental studies are also developing\cite{Bolotin:2008ez,Tombros:2011dh}. 
In addition, there exist various intriguing proposals of the new devices. 
The effects of the top gate on the transport in graphene have been 
studied and several new structures are proposed
\cite{Williams:2011kb,Milovanovic:2013cy, Milovanovic:2013eq, Milovanovic:2013by, Milovanovic:2014fg}. 
A method to control the transport phenomena in graphene using 
strain has been proposed \cite{Pereira:2009ch,Guinea:2012gq,Bahamon:2013ba,Zhang:2014eo} and 
experimental studies are going on\cite{Teague:2009gu,Tomori:2011wl}. 
Together with the nanopatterning of graphene  
\cite{Neumann:2012fw}, 
these kinds of technologies are promising for the future application of graphene. 

In this paper we introduce a new formulation to calculate numerically the electric conductance 
of the mesoscopic conductors within the tight-binding approximation.  
We utilize the so-called singular value decomposition (SVD) in deriving our formalism. 
The SVD is a mathematical technique similar to the matrix diagonalization. 
However, by the SVD, we can treat even the non-square matrices. 
We use the SVD to manipulate the wave-function basis of the conductor. 
Actually, we can omit, using the SVD, some degrees of freedom, which are not
relevant for the transport phenomena, 
thus reducing the dimensions of the matrices in the calculation. 
In addition, this procedure is rather easily parallelized, 
as we will see in Sec. \ref{discussion}. 
Therefore, it is promising for the future application to the 
numerical simulation of the mesoscopic systems in the realistic size. 

On the other hands, a major price we have to pay for the above advantages is that 
our formulation is not useful in treating the electron-electron interactions. 
As one can imagine, once the electron-electron interactions are switched on, 
the omitted degrees of freedom become relevant to the calculation,
thus spoiling the advantages of our formulation. 
We however consider that the large scale calculation of the transport coefficients 
is worth pursuing even without the electron-electron interactions, 
since it gives us the information of the mesoscopic systems, 
which can be directly compared to the experiments. 

The rest of this paper is organized as follows: 
In Sec. \ref{basic}, we describe the basic formalism. 
In Sec. \ref{gen_ide}, we study the ideal wires, 
which are used as the model for the external leads. 
In Sec. \ref{landauer}, we introduce the Landauer formula 
and derive the equation for the electric conductance. 
In Sec. \ref{res-graphene}, we apply our formalism to the 
graphene nanoribbons with external leads. 
In Sec. \ref{discussion}, remaining problems and possible future 
studies are discussed. 
A prospect of the application of parallel computation 
to our formulation is also discussed. 
Supplemental issues are given in Appendices. 

\section{Basic Formalism}
\label{basic}

\subsection{Singular value decomposition}

In developing our formulation, we especially utilized a mathematical technique called 
\lq\lq singular value decomposition (SVD)\rq\rq\cite{Golub:2012wp,Press:2007:NRE:1403886}. 
Using this, we can decompose an arbitrary complex $m\times n$ matrix $A$ as 
\begin{align}
A &= U W V^\dagger
\nonumber\\
&=\left(
\begin{array}{c|c}
\hat{U}&
\tilde{U}
\end{array}
\right)
\left(
\begin{array}{ccc|c}
\sigma_{1} & & \large{O}&\\
& \ddots &&\large{O}\\
\large{O}& &\sigma_{r}&\\\hline
&\large{O}&&\large{O}
\end{array}
\right)
\left(
\begin{array}{c}
\hat{V}^\dagger\\
\hline
\tilde{V}^\dagger
\end{array}
\right),
\label{SVD0}
\end{align}
where $U$ and $V$ are $m\times m$ and $n\times n$ 
unitary matrices, respectively, and 
$\hat{U}$,  
$\tilde{U}$, 
$\hat{V}$ and 
$\tilde{V}$ are block matrices obtained by partitioning $U$ and $V$. 
($X^\dagger$ is the Hermite conjugate of the matrix or vector $X$.)
The integer $r$ is the matrix rank of $A$, {\it i.e.}, $r ={\rm rank} (A)$, and 
$\sigma_j$'s are positive numbers called the singular values of $A$. 
In the following, we write 
$\hat{U} = (\vec{u}_1,\cdots,\vec{u}_r)$,  
$\tilde{U} = (\vec{u}_{r+1},\cdots,\vec{u}_m)$, 
$\hat{V} = (\vec{v}_1,\cdots,\vec{v}_r)$ and 
$\tilde{V} = (\vec{v}_{r+1},\cdots,\vec{v}_n)$, 
where $\vec{u}_j$'s and $\vec{v}_j$'s, respectively, are 
$m$ and $n$ dimensional columnar vectors. 

The following mathematical features are easily confirmed: 
We define the range of $A$, ${\rm Ran}(A)$, and 
the null space of $A$, ${\rm Null}(A)$, as 
\begin{align*}
{\rm Ran}(A) &= \{\vec{y}\in {\mathbb C}^{m}: \vec{y}=A
 \vec{x}\,\,{\text{for some}}\,\,\vec{x}\in {\mathbb C}^{n}\},\\
{\rm Null}(A) &= \{\vec{y}\in {\mathbb C}^{m}: 
A \vec{y}=0\},
\end{align*}
where ${\mathbb C}$ is the set of complex numbers. 
Then, ${\rm Ran}(A)$ is spanned by $\{\vec{u}_1,\cdots,\vec{u}_r\}$ and 
${\rm Null}(A)$ is spanned by $\{\vec{v}_{r+1},\cdots,\vec{v}_n\}$, 
as one can see from Eq. (\ref{SVD0}). 
It is also useful to see that the left null space of $A$, 
namely a set of vectors $\vec{u}$ satisfying 
$\vec{u}^{\,\dagger} A  = 0$, 
is spanned by $\{\vec{u}_{r+1},\cdots,\vec{u}_m\}$, 
which is equivalent to ${\tilde U}^{\,\dagger} A = 0$. 
For convenience, we define an operator, which extract 
$\tilde{U}$ from the SVD of the matrix $A$, {\it i.e.}, Eq. (\ref{SVD0}), as, 
\begin{align}
\tilde{U} = {\cal N}_L[A]. 
\label{tilde}
\end{align}

The most widely known application of the SVD may be  
the so-called Moore-Penrose quasi-inverse matrix, 
which we denote by $A^{\rm MP}$ in this paper. 
The matrix $A^{\rm MP}$ is given by 
\begin{align}
A^{\rm MP} &=\hat{V}
\left(
\begin{array}{ccc}
\sigma_{1}^{-1} & & \large{O}\\
& \ddots &\\
\large{O}& &\sigma_{r}^{-1}
\end{array}
\right)
\hat{U}^\dagger.
\label{MPinverse}
\end{align}
Using this, we can obtain an approximate solution of the equation 
$A\vec{x} =\vec{y}$ as $\vec{x}=A^{\rm MP}\vec{y}$. 
If $\vec{y}\in {\rm Ran}(A)$, 
this solution is exact (but not unique since 
we have freedom to 
add arbitrary superposition of $\vec{v}_{r+1},\cdots,\vec{v}_n$ to $\vec{x}$). 
Useful numerical algorithm to calculate SVD is described in 
Ref. \citen{Press:2007:NRE:1403886}. 

\subsection{Schr\"odinger equation in a recurrence formula}

Here we introduce our model of the mesoscopic conductor. 
The system is described by the tight-binding Hamiltonian, 
\begin{align}
H=-\sum_{\langle m,n\rangle} t_{mn}c_m^\dagger c_n + {\rm H. c.},
\label{Seq_orig}
\end{align}
where $c_m$ is the annihilation operator of an electron 
at the $m$-th site and the summation $\langle m,n\rangle$ is over all the bonds. 
We suppress the spin index throughout this paper. 

From the Heisenberg form, 
\begin{align}
i \hbar \frac{\partial c_m}{\partial t}
=[H,c_m],
\end{align}
the probability conservation relation is expressed as, 
\begin{align}
\frac{\partial \rho_m}{\partial t} =& 
\frac{\partial }{\partial t}c_m^\dagger c_m 
=
\sum_{n\in N(m)}\frac{i}{\hbar}\left(
t_{mn} c_m^\dagger c_n - t_{mn}^* c_n^\dagger c_m
\right)
\nonumber\\
\equiv& 
\sum_{n\in N(m)}J_{nm},
\label{currentop}
\end{align}
where $\rho_m$ is the probability density at the $m$-th site, 
$N(m)$ is the set of sites connected to the $m$-th site and 
$J_{nm}$ stands for the probability current flowing from the site $n$ to $m$. 
We write the field operator of the electron as 
\begin{align}
\Psi = \sum_m \psi_m c_m,
\label{wfun}
\end{align}
and, then, $\psi_m$ is the wave function of the system. 

\begin{figure}[h]
\begin{centering}
\includegraphics[clip,width=7cm]{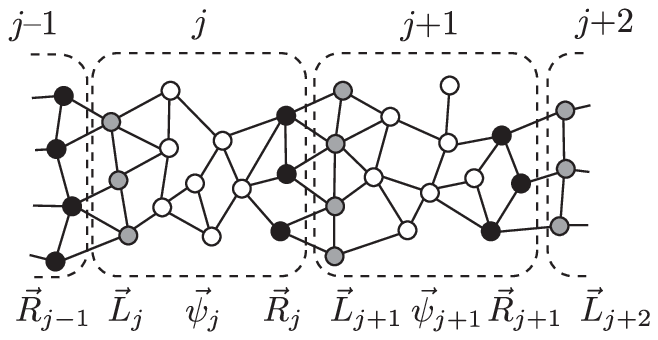} 
\par\end{centering}
\caption{The $(j-1)$-th, $\cdots$, $(j+2)$-th block of the conductor are shown. 
Gray, black and white circles correspond to 
$\vec{R}_j$, $\vec{L}_j$ and $\vec{\psi}_j$, respectively. }
\label{blocks} %
\end{figure}

An example of the conductor is depicted in Fig. \ref{blocks}. 
Each circle corresponds to an atomic site of the tight binding model. 
The whole system is decomposed into several blocks, 
numbered by $j-1$, $j$, $j+1$ and $j+2$ in the figure. 
We divide the wave function $\psi_m$ of the $j$-th block 
into three components, $\vec{L}_j$, $\vec{R}_j$ and $\vec{\psi}_j$, 
each of which is defined as a columnar vector. 
The vector $\vec{L}_j$ ($\vec{R}_j$) represents the wave function of the sites shown by 
gray (black) circles in Fig. \ref{blocks}, 
which are located at the left (right) edge of the $j$-th block 
and are connected by the bonds to the sites in the $(j-1)$-th ($(j+1)$-th) block. 
The vector $\vec{\psi}_j$ represents the wave function of 
the sites in the $j$-th block, which are not included in either $\vec{L}_j$ or $\vec{R}_j$. 
Here we assume that there is no overlap in the elements of $\vec{R}_j$ and $\vec{L}_j$. 
If some elements are overlapped, we can extend the area of the $j$-th block until 
the overlap is dissolved. 
Such extension is always possible unless we introduce the long-range hopping. 

Our aim is to re-express the Schr\"odinger equation of the whole system 
into a set of equations, 
which relate $\{\vec{L}_{j},\vec{R}_{j-1}\}$ to $\{\vec{L}_{j+1},\vec{R}_{j}\}$. 
In order to carry this out, the vectors, $\vec{\psi}_j$'s, should be 
truncated out (or integrated out) by some means. 
We will show that this process can be readily performed 
by employing the SVD and the Moore-Penrose quasi-inverse matrices. 
\cite{Golub:2012wp}

Let us denote the dimensions of $\vec{L}_j$, $\vec{R}_j$ and 
$\vec{\psi}_j$ by $n^L_j$, $n^R_j$ and $n^\psi_j$, respectively, 
and, then, the total number of sites in the $j$-th block is given by 
$n_j = n^L_j+n^R_j+n^\psi_j$. 
The Schr\"odinger equation within the $j$-th block is given by 
\begin{align}
\left(E {I}_{n_j}-{H}_j\right) \cdot
\left(
\begin{array}{c}
\vec{L}_j\\
\vec{R}_j\\
\vec{\psi}_j
\end{array}
\right)=
\left(
\begin{array}{c}
{\Lambda}_j^{LR}\cdot\vec{R}_{j-1}\\
{\Lambda}_j^{RL}\cdot\vec{L}_{j+1}\\
\vec{\rm O}_j
\end{array}
\right),
\label{Seq}
\end{align}
where $E$ is the energy, ${I}_{n}$ is the identity matrix of order $n$ and 
$\vec{\rm O}_j$ is a zero vector whose dimension is $n^{\psi}_j$. 
Here ${H}_j$ is the Hamiltonian of sites within the $j$-th block, 
and ${\Lambda}_j^{RL}$ (${\Lambda}_j^{LR}$) is the hopping matrix elements 
between the $(j-1)$-th and the $j$-th (the $j$-th and the $(j+1)$-th) block. 
Since the Hamiltonian is Hermitian, 
the relation, ${\Lambda}_{j}^{LR}=\{\Lambda_{j-1}^{RL}\}^\dagger$, holds. 
The current flowing from the $(j-1)$-th block to the $j$-th block is given by 
\begin{align}
J &= \frac{i}{\hbar}
\left(\vec{R}^\dagger_{j-1} \vec{L}^\dagger_j\right) 
\left(
\begin{array}{cc}
O& -\Lambda_j^{LR\dagger}\\
\Lambda_j^{LR} & O
\end{array}
\right)
\left(
\begin{array}{c}
\vec{R}_{j-1}\\
 \vec{L}_j
 \end{array}
 \right)
 \nonumber\\
 &\equiv \frac{i}{\hbar}
 \left(\vec{R}^\dagger_{j-1} \vec{L}^\dagger_j\right) 
\Lambda_j
\left(
\begin{array}{c}
\vec{R}_{j-1}\\
 \vec{L}_j
 \end{array}
 \right),
 \label{currentfml}
\end{align}
where we have introduced the matrix $\Lambda_j$. 
Since, in the equilibrium, the probability current conserves at all the boundaries between the blocks, 
the current $J$ does not depend on $j$. 

Rewriting the Eq. (\ref{Seq}), 
so that only $\vec{\psi}_j$ is on the left hand side, yields 
\begin{align}
{K}_j^\psi \vec{\psi}_j &=-{K}_j^L \vec{L}_j -{K}_j^R \vec{R}_j+
\left(
\begin{array}{c}
{\Lambda}_j^{LR}\cdot\vec{R}_{j-1}\\
{\Lambda}_j^{RL}\cdot\vec{L}_{j+1}\\
\vec{\rm O}_j
\end{array}
\right) \equiv\vec{\Gamma}_j,
\label{Seq2}
\end{align}
where ${K}_j^L$, ${K}_j^R$ and ${K}_j^\psi$ are block matrices composing 
the matrix $E{I}_{n_j}-{H}_j = ({K}_j^L \,|\, {K}_j^R \,|\, {K}_j^\psi)$, 
whose dimensions are $n_j\times n_j^L$, $n_j\times n_j^R$ and $n_j\times n_j^\psi$, 
respectively. 
The solvability condition of the Eq. (\ref{Seq2}) 
with respect to $\vec{\psi}_j$ is expressed as 
\begin{align}
\vec{\Gamma}_j \in {\rm Ran}(K_j^\psi). 
\label{conditionG}
\end{align}
From this, the relation between 
$(\vec{R}_{j-1}, \vec{L}_{j})$ and $(\vec{R}_{j}, \vec{L}_{j+1})$ 
is obtained in the following way: 
The Eq. (\ref{conditionG}) means that the vector $\vec{\Gamma}_j$ 
does not belong to the left null space of $K_j^\psi$. 
This is expressed, by introducing $\tilde{U}_j = {\cal N}_L[K_j^\psi]$ 
(see Eq. (\ref{tilde}) for the definition), as 
\begin{align}
\tilde{U}_j^\dagger \vec{\Gamma}_j&=\tilde{U}_j^\dagger\left\{
-{K}_j^L \vec{L}_j -{K}_j^R \vec{R}_j+
\left(
\begin{array}{c}
{\Lambda}_j^{LR}\cdot\vec{R}_{j-1}\\
{\Lambda}_j^{RL}\cdot\vec{L}_{j+1}\\
\vec{\rm O}_j
\end{array}
\right)\right\}
=0.
\label{Seq4}
\end{align}
Conversely, if this holds, we obtain the solution for $\vec{\psi}_j$ as 
$\vec{\psi}_j = \{{K}_j^\psi\}^{\rm MP} \vec{\Gamma}_j$. 
(Note that this is not the unique solution.) 

Introducing the partitioning of $\tilde{U}_j$, ${K}^L_j$ and ${K}^R_j$  
(into upper $n_j^L$ rows, middle $n_j^R$ rows, and 
lower $n_j^\psi$ rows) as 
\begin{align}
\tilde{U}_j = \left(
\begin{array}{c}
\tilde{U}^L_j\\
\tilde{U}^R_j\\
\tilde{U}^\psi_j
\end{array}
\right),\,\,
{K}^L_j = \left(
\begin{array}{c}
{K}^{LL}_j\\
{K}^{RL}_j\\
{K}^{\psi L}_j
\end{array}
\right),\,\,
{K}^R_j = \left(
\begin{array}{c}
{K}^{LR}_j\\
{K}^{RR}_j\\
{K}^{\psi R}_j
\end{array}
\right),
\end{align}
the Eq. (\ref{Seq4}) is rewritten as 
\begin{align}
-&A_j\vec{L}_j-B_j\vec{R}_j+\tilde{U}_j^{L\dagger} \Lambda_j^{LR}\vec{R}_{j-1}
+\tilde{U}_j^{R\dagger} \Lambda_j^{RL}\vec{L}_{j+1}=0,
\end{align}
where 
\begin{align}
A_j=\tilde{U}_j^{L\dagger} K^{LL}_j
+\tilde{U}_j^{R\dagger} {K}^{RL}_j+\tilde{U}_j^{\psi\dagger} K^{\psi L}_j,
\nonumber\\
B_j=\tilde{U}_j^{L\dagger} {K}^{LR}_j+\tilde{U}_j^{R\dagger} K^{RR}_j
+\tilde{U}_j^{\psi\dagger} K^{\psi R}_j.
\end{align}
Rearranging these equations, we obtain the relation between $(\vec{R}_{j-1}, \vec{L}_j)$ and
$(\vec{R}_{j}, \vec{L}_{j+1})$ as
\begin{align}
&{P}_j
\left(
\begin{array}{c}
\vec{R}_{j-1}\\
\vec{L}_j
\end{array}
\right)
=
{Q}_j 
\left(
\begin{array}{c}
\vec{R}_j\\
\vec{L}_{j+1}
\end{array}
\right)
\label{transfer}
\end{align}
where 
\begin{align}
{P}_j &= \left(
\begin{array}
{c|c}
\tilde{U}_j^{L\dagger}\Lambda_j^{LR}  &
-A_j
\end{array}
\right),
\nonumber\\
{Q}_j &= 
\left(
\begin{array}
{c|c}
B_j &
-\tilde{U}_j^{R\dagger} \Lambda_j^{RL}
\end{array}
\right).
\end{align}

In case of $n_j^\psi = 0$, 
we should put 
$\tilde{U}_j = I_{n_j^R + n_j^L}$ and omit 
all the terms with $\psi$ symbol. 

\subsection{Boundary condition and transfer matrix}

Let us write as 
$\displaystyle 
\vec{\Phi}_j =\left(
\begin{array}{c}
\vec{R}_{j-1}\\
\vec{L}_j
\end{array}
\right) $.
When the whole system is composed of $M$ blocks, 
the transport propertied of the system is described by 
the vectors, $\vec{\Phi}_1,  \cdots, \vec{\Phi}_{M+1}$. 
The vector $\vec{\Phi}_1$ includes $\vec{R}_0$ and 
$\vec{\Phi}_{M+1}$ includes $\vec{L}_{M+1}$. 
The wave functions $\vec{R}_0$ and $\vec{L}_{M+1}$ are 
located out of the $M$ blocks and 
we assume that these give the boundary conditions 
for the conductor. 

Our next goal is to derive the equation which directly relates 
the components, $\vec{\Phi}_1$ and $\vec{\Phi}_{M+1}$. 
We rewrite whole equations into the form ${A} \vec{\Phi} = \vec{\Theta}$ where 
\begin{align}
&{A}=\left(
\begin{array}{cccc}
 {Q}_1 &&&\\
{P}_2 & {Q}_2 &&{O}\\
&\ddots & \ddots &\\
&&{P}_{M-1} & {Q}_{M-1} \\
&{O}&&{P}_M \\
\end{array}
\right),
\nonumber\\
&\vec{\Phi}=\left(
\begin{array}{c}
\vec{\Phi}_2\\
\vec{\Phi}_3\\
\vdots\\
\vec{\Phi}_{M-1}\\
\vec{\Phi}_{M}
\end{array}
\right),\,\,
\vec{\Theta}=
\left(
\begin{array}{c}
-{P}_1\vec{\Phi}_1\\
0\\
\vdots\\
0\\
-{Q}_M\vec{\Phi}_{M+1}
\end{array}
\right).
\label{Seq5}
\end{align}
As discussed in the previous section, the solvability condition of 
${A} \vec{\Phi} = \vec{\Theta}$ with respect to 
$\vec{\Phi}$ is given by $\vec{\Theta} \in {\rm Ran}(A)$, 
from which we can derive an equation satisfied by $\vec{\Theta}$. 

We can see that $\vec{\Theta}$ satisfies the condition $\tilde{U}^\dagger \vec{\Theta} =0$, 
where $\tilde{U} = {\cal N}_L[A]$. 
We introduce the partitioning $\tilde{U}^\dagger = (\tilde{U}_1^{A\dagger} |\cdots |\tilde{U}^{A\dagger}_{M+1})$, 
where $\tilde{U}^A_1$ and $\tilde{U}^A_{M+1}$ are block matrices, 
whose columns correspond respectively to the components, 
$-P_1 \vec{\Phi}_1$ and $-Q_M \vec{\Phi}_{M+1}$, of $\vec{\Theta}$ 
in Eq. (\ref{Seq5}). 
Then, we obtain the relation between $\vec{\Phi}_1$ and $\vec{\Phi}_{M+1}$ as 
\begin{align}
\tilde{U}^{A\dagger}_1 {P}_1\vec{\Phi}_1 
= -\tilde{U}^{A\dagger}_M {Q}_M\vec{\Phi}_{M+1}.
\end{align} 
Denoting $\tilde{U}^{A\dagger}_1 {P}_1$ and 
$-\tilde{U}^{A\dagger}_M {Q}_M$ by ${P}^A$ and ${Q}^A$, respectively, 
and introducing the wave functions of the left lead 
$\vec{\Phi}^l=\vec{\Phi}_1$ and the 
right lead $\vec{\Phi}^r=\vec{\Phi}_{M+1}$,
we obtain the transfer matrices which connect the left and the right leads as 
\begin{align}
{P}^A \vec{\Phi}^l = {Q}^A\vec{\Phi}^r. 
\label{lr-relation}
\end{align}
From this we can calculate the scattering matrix of the conductor. 

\section{A Generalized Ideal Wire}
\label{gen_ide}

Here we study the ideal wires as a model for the leads. 
We assume that the wire is made of a periodic reputation of the 
set of atoms. 
In that case, all the matrices ${P}_j$ (${Q}_j$) do not depend 
on $j$ and we denote them simply by ${P}$ (${Q}$). 

We assume that the eigenstates of the lead satisfy the relation \cite{Ando:1991wt}
\begin{align}
\vec{\Phi}_j = z \vec{\Phi}_{j-1} ,
 \label{wf1}
 \end{align}
where $z$ is a complex number. 
From the Eq. (\ref{transfer}), 
the equation determining $z$ and the corresponding eigenvector is given by
\begin{align}
{P}\vec{\Phi}_j 
 =z{Q}\vec{\Phi}_j.
 \label{pqeq}
 \end{align}
Here we call $z$ the {\it transfer eigenvalue} of the wire and the 
corresponding eignevector, $
\vec{\phi}_z$,
the {\it transfer eigenvector} (or {\it eigenstate}). 
We should note that $P$ and $Q$ are not necessarily square matrices. 
In that sense, this is a generalized eigenvalue problem. 
A method to solve this equation is given in the Appendix\ref{GEP}. 

There are several types of eigenmodes: 
the mode with $|z|=1$ corresponds to 
a propagating mode (or channel), and that with $|z|\ne 1$ 
to an evanescent mode. 
When $|z|>1$ ($|z|<1$) the wave function grows (decays) 
as $j$ increases. 
Some evanescent modes 
also have imaginary part in $z$ and 
they oscillate as well as grow or decay. 

Here we may assume the following mathematical features: 
\begin{enumerate}
\item[1)] If $z$ is a transfer eigenvalue, $1/z$ also is. 
In case of $|z|=1$, $1/z = z^*$. 
\item[2)] The current matrix $\Lambda_j$ (see Eq. (\ref{currentfml})) is
diagonalized as 
\begin{align}
{}^t \vec{\phi}_{1/z}\Lambda_j \vec{\phi}_{z' }=0 \,\,\,\,
(\text{if}\, z\ne z')
\label{current-diagonal}
\end{align}
by the transfer eigenstates. 
(Some note is given in the Appendix \ref{degeneracy}.)
\end{enumerate}
Here we note that 1) seems clear if the system has a time-reversal symmetry. 
However, similar feature may also exist even without time-reversal symmetry. 

In the preceding works, it has sometimes been assumed that 
the direction parallel and perpendicular to the wire are separable, 
such as the case of a simple wire made of a square lattice. 
In that case, we can define the transverse modes rather easily and 
the property expressed in 2) is clearly satisfied. 
Our statement is the generalization of such a case. 
It seems that 2) is quite reasonable from the physical point of view 
even in generalized situations, although we do not go 
into details in this paper.

\section{Landauer Formula and Conductance}
\label{landauer}

\begin{figure}[h]
\begin{centering}
\includegraphics[clip,width=7.8cm]{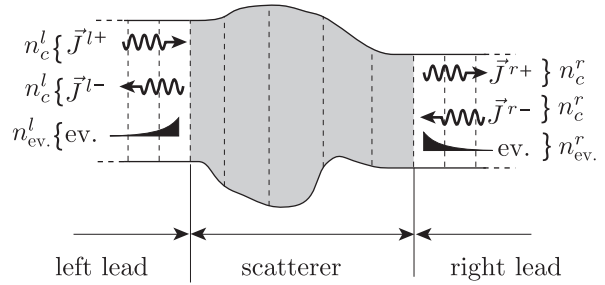} 
\caption{Geometry of the conductor is schematically shown. 
The shaded area is the scatterer connected to the left and the right leads. 
Vertical dashed lines are boundaries of the blocks. 
Symbols with \lq\lq ev.\rq\rq\, indicate 
the evanescent modes. 
The numbers of each modes are also indicated.}
\label{scatterer} %
\end{centering}
\end{figure}

We consider the situation depicted in Fig. \ref{scatterer}. 
The scatterer is connected to the left and the right leads. 
Two leads are not necessarily identical to each other. 
However each of them should be have a completely periodic structure, 
so that the channels are well-defined. 
We assume that the two leads are semi-infinitely long. 

As depicted in Fig.\ref{scatterer}, we consider 
the right- and the left-moving modes 
and the evanescent modes decaying with distance from the scatterer. 
We assume that there are $n_c^l$ ($n_c^r$) channels and 
$n_{\rm ev.}^l$ ($n_{\rm ev.}^r$) evanescent modes in 
the left (right) lead. 
(Note that one channel corresponds to a pair of the right- and the left-moving mode.) 

Now we calculate the conductance of this conductor. 
To do this, we utilize the so-called Landauer formula. 
First we introduce $(n^l_c+n^r_c)\times (n^l_c+n^r_c)$ matrix $S$, 
which relates the probability currents of the incoming waves to that of the outgoing waves, as
\begin{align}
\vec{J}^{\,\,{\rm out}}\equiv \left(
\begin{array}{c}
\vec{J}^{\,l-}\\
\vec{J}^{\,r+}
\end{array}
\right)
=\left(
\begin{array}{cc}
R & T'\\
T & R'
\end{array}
\right)
 \left(
\begin{array}{c}
\vec{J}^{\,l+}\\
\vec{J}^{\,r-}
\end{array}
\right)
\equiv S \vec{J}^{\,\,{\rm in}},
\end{align}
where $\vec{J}^{\,l+}$, etc. 
represent the probability currents carried by the propagating modes; 
\lq\lq$l$\,\rq\rq\, and \lq\lq$r\,$\rq\rq\, indicate respectively the left and the right lead, and 
\lq\lq$+$\rq\rq\, and \lq\lq$-$\rq\rq\, the right-moving and the left-moving mode. 

We denote the wave functions at the left and the right lead 
by $\vec{\Phi}^l$ and $\vec{\Phi}^r$, respectively. 
Each function is given as a superposition of the right-moving, 
the left-moving and the evanescent modes as 
\begin{align}
\vec{\Phi}^l &= \sum_{j=1}^{n^l}
\left(
\alpha_j^{l+} \vec{\phi}_j^{\,l+}+ 
\alpha_j^{l-} \vec{\phi}_j^{\,l-} 
\right)
+\sum_{j=1}^{n_{\rm ev}^l}
\alpha_j^{l<}\vec{\phi}^{\,l<}_j,
\nonumber\\
\vec{\Phi}^r &= \sum_{j=1}^{n^r}
\left(
\alpha_j^{r+} \vec{\phi}_j^{\,r+}+ 
\alpha_j^{r-} \vec{\phi}_j^{\,r-} 
\right)
+\sum_{j=1}^{n_{\rm ev}^r}
\alpha_r^{l>}\vec{\phi}^{\,r>}_j,
\label{waves}
\end{align}
where $\alpha_j^{l(r)\pm}$ and $\alpha_j^{l<(r>)}$ are constants. 
The vectors $\vec{\phi}_j^{\,l(r)+}$ and $\vec{\phi}_j^{\,l(r)-}$ represent 
the eigenvectors of the right-moving and the left-moving modes, respectively, 
in the left (right) lead. 
The vector $\vec{\phi}_j^{\,l<(r>)}$ represents the wave function of 
the evanescent modes in the left (right) lead.

Here we introduce the matrix representation of a basis of the wave function as 
$\Theta^{l+} = (\vec{\phi}_1^{\,l+}\cdots,\vec{\phi}_{n^l}^{\,l+})$. 
Then the Eqs. (\ref{waves}) are rewritten as
\begin{align}
\vec{\Phi}^l &= \Theta^{l+}\vec{\alpha}^{\,l+} +\Theta^{l-}\vec{\alpha}^{\,l-} 
+\Theta^{l<}\vec{\alpha}^{\,l<},
\nonumber\\
\vec{\Phi}^r &= \Theta^{r+}\vec{\alpha}^{\,r+} +\Theta^{r-}\vec{\alpha}^{\,r-} 
+\Theta^{r>}\vec{\alpha}^{\,r>},
\label{waves2}
\end{align}
where $\vec{\alpha}^{l+}_j = {}^t(\alpha_1^{l+},\cdots,\alpha_{n^l}^{l+})$, etc. 
are columnar vectors. 
From the Eq. (\ref{lr-relation}), we obtain 
\begin{align}
&P^A \left(\Theta^{l+}\vec{\alpha}^{\,l+} +\Theta^{l-}\vec{\alpha}^{\,l-} 
+\Theta^{l<}\vec{\alpha}^{\,l<}\right)  \nonumber\\
&=
Q^A \left(\Theta^{r+}\vec{\alpha}^{\,r+} +\Theta^{r-}\vec{\alpha}^{\,r-} 
+\Theta^{r>}\vec{\alpha}^{\,r>}
\right).
\end{align}
This can be rewritten in a matrix form as
\begin{align}
&F
\left(
\begin{array}{c}
\vec{\alpha}^{\,l+} \\
\vec{\alpha}^{\,r-}\\
\vec{\alpha}^{\,l-} \\
\vec{\alpha}^{\,r+}
\end{array}
\right)=
G\left(
\begin{array}{c}
\vec{\alpha}^{\,l<}\\
\vec{\alpha}^{\,r>} 
\end{array}
\right),
\label{eq29}
\end{align}
where 
\begin{align}
F&= \left(
\begin{array}{c|c|c|c}
P^A\Theta^{l+}& 
-Q^A\Theta^{r-}&
 P^A\Theta^{l-}& 
 -Q^A\Theta^{r+}
\end{array}
\right),
\nonumber\\
G&= \left(
\begin{array}{c|c}
-P^A\Theta^{l<}& 
Q^A\Theta^{r>}
\end{array}
\right).
\end{align}
Here we introduce $\tilde{U}_G = {\cal N}_L[G]$ 
(see Eq. (\ref{tilde})). 
Then, the solvability condition of the Eq. (\ref{eq29}) 
with respect to $\vec{\alpha}^{\,l<}$ and 
$\vec{\alpha}^{\,r>}$ is given by 
\begin{align}
\tilde{U}_G^\dagger F \left(
\begin{array}{c}
\vec{\alpha}^{\,l+} \\
\vec{\alpha}^{\,r-}\\
\vec{\alpha}^{\,l-} \\
\vec{\alpha}^{\,r+}
\end{array}
\right) 
= 0,
\label{nlg}
\end{align}
which relates the amplitudes of the incoming and the outgoing waves. 
Note that, if there are no evanescent modes, $\tilde{U}_G$
should be the identity matrix of order $2 (n_c^l+n_c^r)$. 

Writing $\tilde{U}_G^\dagger F = (D^{\rm in}\,|\,D^{\rm out})$, 
where $D^{\rm in(out)}$ is the left (right) $n_c^l+n_c^r$ columns 
of $\tilde{U}_G^\dagger F$, 
and introducing $\displaystyle \vec{\alpha}^{\rm in} \equiv{}
\left(
\begin{array}{c}
\vec{\alpha}^{l+}\\
\vec{\alpha}^{r-}
\end{array}
\right)$, 
$\displaystyle \vec{\alpha}^{\rm out} \equiv{}
\left(
\begin{array}{c}
\vec{\alpha}^{l-}\\
\vec{\alpha}^{r+}
\end{array}
\right)$, 
Eq. (\ref{nlg}) becomes
$D^{\rm out}\vec{\alpha}^{\rm out} = 
-D^{\rm in}
\vec{\alpha}^{\rm in}$. 
Using the Moore-Penrose inverse of $D^{\rm out}$ 
and defining $D^{\rm tot}\equiv \{D^{\rm out}\}^{\rm MP}D^{\rm in}$, 
we reach the expression
\begin{align}
\vec{\alpha}^{\rm out} = 
-D^{\rm tot}
\vec{\alpha}^{\rm in}.
\label{alpha1}
\end{align}

From the argument of the Sec. \ref{gen_ide} and the Eq. (\ref{current-diagonal}), 
the total probability current is given by the sum of the 
independent contributions of each channel. 
Let us denote the contribution of the $j$-th \lq\lq in (out)\rq\rq\, 
channel by $\eta^{\rm in (out)}_j$, 
the $j$-th components of $\vec{J}^{\,\,{\rm in}}$ and $\vec{J}^{\,\,{\rm out}}$ are given by 
\begin{align}
J^{\,\,{\rm in}}_j &= \eta^{\rm in}_j |\alpha_j^{\,{\rm in}}|^2,\,\,
J^{\,\,{\rm out}}_j = \eta^{\rm out}_j |\alpha_j^{\,{\rm out}}|^2.\,\,
\nonumber
\end{align}
Using Eq. (\ref{alpha1}), 
the probability current carried by the $j$-th outgoing channel is given by
\begin{align}
J_j^{\rm out} &= \sum_{k,k'=1}^{n^l_c+n^r_c} D^*_{jk}
\eta_j^{\rm out} D^{\rm tot}_{jk'}\left(\alpha^{\rm in}_k\right)^* \alpha^{\rm in}_{k'}
\nonumber\\
&= \sum_{k=1}^{n^l_c+n^r_c} 
\eta_j^{\rm out} \left|D^{\rm tot}_{jk}\right|^2
\left|\alpha^{\rm in}_k\right|^2
\nonumber\\
&\phantom{=}+
\sum_{k,k'=1 (k \ne k')}^{n^l_c+n^r_c} D^{\rm tot\,*}_{jk}
\eta_j^{\rm out} D^{\rm tot}_{jk'}\left(\alpha^{\rm in}_k\right)^* \alpha^{\rm in}_{k'}.
\label{jout}
\end{align}
Here we separated the summation into the diagonal and 
the off-diagonal terms with respect to $k$ and $k'$. 
We assume that electrons are injected to 
the modes $\alpha^{\rm in}_k$ from the reservoir incoherently, 
and then the cross term (the last line of Eq. (\ref{jout})) 
vanishes after time averaging. 
As a result we obtain the following relation, 
\begin{align}
J_j^{\rm out} &= \sum_{k,=1}^{n^l+n^r_c} 
\eta_j^{\rm out} \left|D^{\rm tot}_{jk}\right|^2\left|\alpha^{\rm in}_k\right|^2
=\sum_{k,=1}^{n^l_c+n^r_c} 
S_{jk}J_k^{\rm in}.
\label{jout2}
\end{align}
Then the elements of the scattering matrix $S$ is given by 
\begin{align}
S_{jk} = \frac{\eta_j^{\rm out} }{\eta_k^{\rm in}}\left|D^{\rm tot}_{jk}\right|^2.
\end{align}

Using the Landauer formula we can calculate the conductance as \cite{Buttiker:1985vk}
\begin{align}
G = \frac{2 e^2}{h}\sum_{j=n^l_c+1}^{n^r_c+n^l_c}
\sum_{k=1}^{n^l_c} S_{jk}. 
\end{align}

In actual calculations, the wave functions should be normalized appropriately. 
However, the normalization is not relevant to the results, 
since the present formulation depends only on 
the ratio of the incoming and the outgoing current. 
Therefore we can adopt any normalization for our convenience. 

\section{Application to Graphene Constrictions}
\label{res-graphene}

We apply the present formulation to the graphene constrictions of the several forms. 
The character of the graphene is taken into account by assuming the 
honeycomb lattice structure. 
The hopping is restricted to the nearest neighbors for simplicity. 

\begin{figure}[h]
\begin{centering}
\includegraphics[clip,width=7cm]{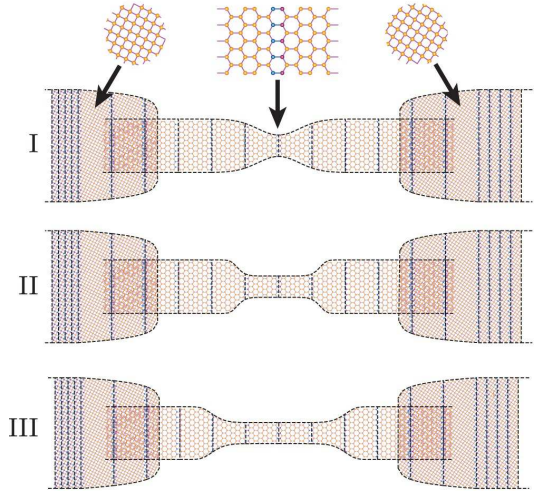} 
\par\end{centering}
\caption{The geometries of the graphene nanoribbons 
studied in this paper are shown. 
I, II and II have constrictions in the middle of the conductor, 
though the lengths of the constricted regions differ. 
The constricted region is $N=4$ armchair nanoribbon 
in all cases as one can see from the magnification. 
Both leads are modeled by square lattices. 
}
\label{Geometry2} %
\end{figure}

\begin{figure}[h]
\begin{centering}
\includegraphics[clip,width=6.5cm]{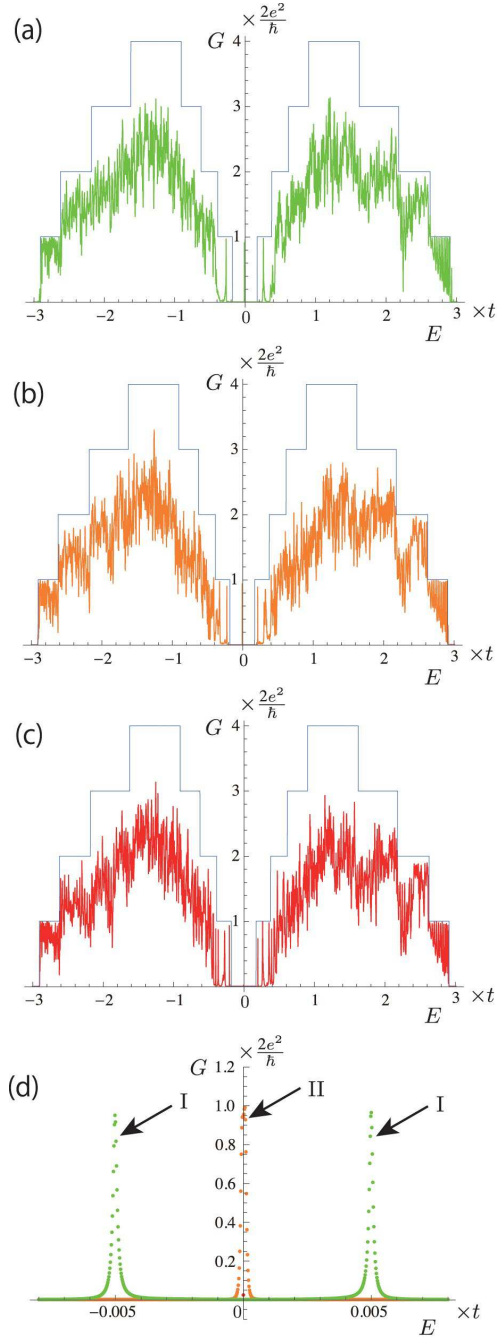} 
\par\end{centering}
\caption{The energy dependence of the conductance of the wires 
depicted in Fig. \ref{Geometry2}. 
(a), (b) and (c) correspond to 
I, II, and III of Fig. \ref{Geometry2}, respectively. 
The vertical axis is the conductance $G$ including spin degree of freedom. 
The step-like function, shown in three figures, is the conductance of an ideal $N=4$ armchair wire. 
(d) is the magnification near the charge neutrality point $E=0$. 
The peaks are those of I and II, as shown in the figure. 
III also shows a small peak at $E=0$, however it is non-distinctive. }
\label{conductance} %
\end{figure}

We treat the graphene wires with a short (I), a middle (II), and a long (III) constriction 
as depicted in Fig. \ref{Geometry2}. 
The narrowest parts of the wires are composed of $N=4$ armchair nanoribbons.  
The left and the right leads are assumed to be made of 
two-dimensional square lattices laid parallel to the graphene surface, 
whose lattice spacing is $0.55\times a$ ($a$: the minimum C-C spacing in the graphene). 
The directions of the square lattices are intentionally rotated 
from the wire axis. 
The transfer integrals within the leads ($t_{\rm lead}$) and those 
between the graphene and the leads ($t'$) are set as $t_{\rm lead}=t$ and $t'=0.3\times t$.  
Only the nearest neighbor hopping is assumed within the leads, and 
the hopping between the leads and the graphene is assumed only 
between the sites, whose horizontal separation (parallel to the graphene surface) 
is smaller than $2\times a$. 

The conductance of the wires at zero temperature are calculated as a function of the energy and the 
results are shown in Fig. \ref{conductance} (a) $\sim$ (c). 
The apparent step like function (indicated in blue) is the conductance of an ideal $N=4$ armchair wire. 
As one can see, the conductance curves of the constrictions consist of spiky peaks, 
which may arise from the various resonances of the conduction electrons. 
The resemblance to the blue steps are not obvious in all cases. 
However at some energies the conductance approaches 
the perfect transmission, especially in the lowest step $G= 2 e^2/h$. 

Near zero energy, some peaks are seen in (a) and (b) 
(The magnification is shown in (d)). 
These may come from the transmission through the evanescent 
modes decaying in the constricted region. 
This explanation is plausible since such peaks are not significant in 
the case of the longest constriction (c), 
though a tiny peak still exists. 

\begin{figure}[h]
\begin{centering}
\includegraphics[clip,width=6.5cm]{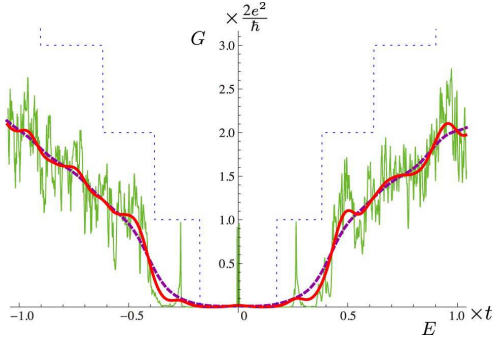} 
\par\end{centering}
\caption{Conductance of Fig.\ref{Geometry2} (a) after the thermal average. 
The thin green curve, thick red curve and dashed purple curve 
show the result of $T=0$, $T=0.05\times t$ and $T=0.1\times t$, 
respectively. }
\label{thermal} %
\end{figure}

The conductance at a finite temperature is calculated from
\begin{align}
G (E,T) &= \int_{-\infty}^\infty G(E',0) f'(E'-E) dE'
\nonumber\\
&=\int_{-\infty}^\infty G(E',0)\frac{1}{4 T}\frac{1}{\cosh^2 ((E'-E)/(2T))} dE'
\end{align}
where $f(E)$ is the Fermi distribution function and 
$G(E,T)$ is the conductance at the bias $E$ and the temperature $T$. 
In Fig. \ref{thermal}, we have shown the results of the constriction I for 
$T=0, 0.05\times t$ and $0.1\times t$. 
At $T=0.05\times t$, we can see several step-like structures, 
however their relation to the steps of the ideal $N=4$ wire is not clear at present. 
More intensive study is required to clarify the conductance quantization in 
graphene constrictions. 

We note on the particle-hole symmetry of these systems. 
It has been known that the graphene nanoribbons show particle-hole symmetry 
if the hopping is limited to the nearest neighbors (NN's). 
The same property holds for the square lattices. 
With the particle-hole symmetry, the conductance 
is symmetric with respect to the energy inversion $E\rightarrow -E$. 
The particle-hole symmetry in the graphene 
is broken when the next nearest neighbor (NNN) hopping is switched on. 
In the present calculation, the graphene and 
the leads separately hold particle-hole symmetry. 
However, the coupling of the graphene to the leads 
effectively induces the NNN hopping in the graphene, 
which may probably break the particle-hole symmetry. 
The slightly asymmetric behaviors, as seen in 
Figs. \ref{conductance} and \ref{thermal}, 
may originate from such a mechanism. 

\section{Discussion}
\label{discussion}

In this paper we have introduced a new formulation to calculate 
the conductance of the mesoscopic conductors. 
The key point of our formulation is to omit, using the SVD, the degrees of freedom 
in the scatterer irrelevant for the transport phenomena. 
This procedure may be available for other scattering problems, 
although, until now, the SVD does not seem to be used so often in the studies of physics. 

In the present study we have not treated the 
random potential, 
the magnetic field,
the spin orbit coupling, and 
the lattice distortion.
Inclusion of these into our formalism may be straight forward. 
They will provide useful information in the various fields, 
such as spintronics or topological insulators\cite{Kane:2005gb}. 

It is also important to treat the multi-terminal geometry\cite{Baranger:1991ta}. 
For example, the Hall conductance belongs to this category. 
We consider that the present treatment can be extended 
to such situations by introducing appropriate partitioning of 
the sample, which, however, needs some more theoretical efforts. 

Finally, we point out a possibility of applying 
parallel computation to our formulation. 
Here we show that the dimension of the large matrix 
appearing in the Eq. (\ref{Seq5}) can be reduced 
by the following method. 
Let us take a part of the equation, 
\begin{align}
P_j \vec{\Phi}_j &= Q_j \vec{\Phi}_{j+1},\\
P_{j+1} \vec{\Phi}_{j+1} &= Q_{j+1} \vec{\Phi}_{j+2}.
\end{align}
We rewrite this by introducing block matrices as, 
\begin{align}
\left(
\begin{array}{c|c}
P_j & O\\
\hline
O & Q_{j+1}
\end{array}
\right)
\left(
\begin{array}{c}
\vec{\Phi}_j\\
\hline
\vec{\Phi}_{j+2}
\end{array}
\right)
= 
\left(
\begin{array}{c}
Q_j\\
\hline
P_{j+1}
\end{array}
\right)
\vec{\Phi}_{j+1}.
\end{align}
Then, we obtain the solvability condition for $\vec{\Phi}_{j+1}$ as
\begin{align}
{\cal N}_L
\left[
\left(
\begin{array}{c}
Q_j\\
\hline
P_{j+1}
\end{array}
\right)
\right]^\dagger
\left(
\begin{array}{c|c}
P_j & O\\
\hline
O & Q_{j+1}
\end{array}
\right)
\left(
\begin{array}{c}
\vec{\Phi}_j\\
\hline
\vec{\Phi}_{j+2}
\end{array}
\right)=0.
\end{align}
Rewiring the blocks of the matrix by 
\begin{align}
&P'_j\equiv {\cal N}_L
\left[
\left(
\begin{array}{c}
Q_j\\
\hline
P_{j+1}
\end{array}
\right)
\right]^\dagger
\left(
\begin{array}{c}
P_j\\
\hline
O
\end{array}
\right),
\nonumber\\
&Q'_j\equiv -{\cal N}_L
\left[
\left(
\begin{array}{c}
Q_j\\
\hline
P_{j+1}
\end{array}
\right)
\right]^\dagger
\left(
\begin{array}{c}
O\\
\hline
Q_{j+1}
\end{array}
\right),
\end{align}
we result in 
\begin{align}
P'_j\vec{\Phi}_j= Q'_{j+1}\vec{\Phi}_{j+2}. 
\end{align}
Now $\vec{\Phi}_{j+1}$ is deleted and the matrix dimension is reduced. 

By using the above transformation we can reduce the number of 
blocks in the large matrix in the Eq. (\ref{Seq5}) by one. 
Applying this method repeatedly, the matrix dimension is remarkably reduced. 
Furthermore, this process can be carried out for several different $j$'s 
simultaneously, which allows us to speed up the calculation by using 
parallel computation. 

Using the above procedure, we can enlarge the system size drastically, 
which potentially allows us the real-scale calculations at the atomic level. 
As we have pointed out before, our formulation is not useful 
for the systems with the electron-electron interaction. 
However we consider that the advantages of our formulation 
overcome such a disadvantage. 

\section{Summary}

We have proposed a new formulation to calculate the electric conductance 
of the mesoscopic conductors, 
which has a potential advantage in treating huge systems 
using parallel computation. 
A simple example of the calculation is shown for graphene nanoribbons 
with external leads. 


\vspace{3mm}
\noindent
{\bf Acknowledgement}

The author is grateful to H. Yoshioka, A. Kanda and H. Tomori for 
valuable discussions.  
Especially, he owes the argument on the particle-hole symmetry 
in Sec. \ref{res-graphene} largely to H. Yoshioka. 
 
This work was supported by JSPS KAKENHI Grant Numbers 24540392, 22540329. 


\appendix

\section{A method to solve the generalized eigenvalue problem}
\label{GEP}

We present a simple method to solve 
the equation of the following type,
\begin{align}
P \vec{x} = z Q \vec{x}
\label{geig}
\end{align}
where $P$ and $Q$ are $M\times N$ matrices ($M\ne N$ in general). 
A number, $z$, and a vector, $\vec{x}$, should be determined. 
In solving Eq. (\ref{geig}), it is advantageous to 
formulate the problem on the basis of the eigenvalue problem, 
although some attention is needed. 
The calculation consists of the following two steps. 

\subsection{Removal of the intersection of the null spaces of $P$ and $Q$}

First we note the case where the intersection of 
the null spaces of the matrices ${P}$ and ${Q}$ is not empty. 
Let us suppose that a vector $\vec{v}$ is in the null spaces of ${P}$ and ${Q}$ simultaneously, 
namely, $\vec{v}$ satisfies ${P}\vec{v}=0$ and ${Q}\vec{v}=0$.  
Then, $\vec{v}$ is a solution of Eq. (\ref{geig}) irrespective of the value $z$. 
In such case, we cannot determine the eigenvectors other than $\vec{v}$ from Eq. (\ref{geig}), 
since even if $P$ and $Q$ are square matrices, 
the characteristic polynomial ${\rm det}(P - z Q)$ vanishes for any $z$ and, then, 
the eigenvalues are not determined. 
In order to avoid this situation we need to remove 
the intersection of the null spaces of $P$ and $Q$ from the basis of the vectors. 
This process is executed as follows (see Chap. 12.4 of Ref. \citen{Golub:2012wp}).  

First we consider a block matrix $\Omega$ made from $P$ and $Q$ 
and introduce its SVD as follows , 
\begin{align}
\Omega &\equiv \left(
\begin{array}{c}
P \\ \hline Q
\end{array}
\right)
=
\left(
\tilde{U}\,\,
\hat{U}
\right)
\left(
\begin{array}{ccc|c}
\sigma_{1} & &\large{O} &\\
& \ddots &&\large{O}\\
\large{O}& &\sigma_{r}&\\
\hline
&\large{O}&&\large{O}
\end{array}
\right)
\left(
\begin{array}{c}
\tilde{V}^{\dagger}\\
\hat{V}^{\dagger}
\end{array}
\right)\nonumber\\
&=
\tilde{U}\left(
\begin{array}{ccc}
\sigma_{1} & &\\
& \ddots &\\
& &\sigma_{r}
\end{array}
\right)
\tilde{V}^{\dagger}
\equiv \tilde{U}
\tilde{W}\tilde{V}^{\dagger},
\label{SVDPQ}
\end{align}
where $\sigma_j$'s are the singular values of $\Omega$, and $r={\rm rank}\,(\Omega)$. 
Here the columns of $V=(\tilde{\vec{v}}_1,\cdots,\tilde{\vec{v}}_{N})$ 
form a complete orthonormal basis of the $N$-dimensional space, and  
the columns of $\hat{V}=(\tilde{\vec{v}}_{r+1},\cdots,\tilde{\vec{v}}_{N})$ 
form the basis of the null space of $\Omega$, 
which is actually the intersection of the null spaces of $P$ and $Q$. 
Therefore if we restrict the basis of $\vec{x}$\, to  
$\vec{v}_1\cdots \vec{v}_r$, 
the intersection of the null spaces of $P$ and $Q$ is 
removed from the solution space of $\vec{x}$. 

Here we set
\begin{align}
\vec{x} = \sum_{j=1}^r \frac{y_j}{\sigma_j} \vec{v}_j = \tilde{V}\tilde{W}^{-1}\vec{y}.
\end{align}
If we divide $\tilde{U}$ into upper $M$ rows and lower $M$ rows and put 
\begin{align}
\left(
\begin{array}{c}
P \\ \hline Q
\end{array}
\right)
=\left(
\begin{array}{c}
\Pi \\ \hline \Theta
\end{array}
\right)\tilde{W}\tilde{V}^\dagger, 
\end{align}
the relations $P=\Pi\tilde{W}\tilde{V}^\dagger$ and 
$Q=\Theta\tilde{W}\tilde{V}^\dagger$ hold and 
the Eq. (\ref{geig}) is rewritten as 
\begin{align}
\Pi\,\vec{y} = z \,\Theta\, \vec{y}.
\label{geig2}
\end{align}
Note that $\tilde{V}^\dagger \tilde{V} = I_r$. 
At this stage, $\Pi$ and $\Theta$ are not necessarily 
square matrices.

\subsection{Making matrices $\Pi$ and $\Theta$ square}

Next we make the matrices $\Pi$ and $\Theta$ square 
by truncating some of the null spaces of $\Pi$ or $\Theta$. 
We put the dimensions of $\Pi$ and $\Theta$ 
to be $m\times n$ and also define $d \equiv {\rm min}(m,n)$\cite{truncation}. 
The SVD of $\Pi$ and $\Theta$ are given as 
\begin{align}
\Pi = U^\Pi W^\Pi V^{\Pi\dagger},\,\,\,
\Theta = U^\Theta W^\Theta V^{\Theta\dagger}
\end{align}
and we truncate these matrices as in the same manner as 
the Eq. (\ref{SVDPQ}) so that $W^\Pi$ and $W^\Theta$ become $d\times d$ matrices. 
The truncated matrices are indicated by tilde signs. 
Then the Eq. (\ref{geig2}) is rewritten as 
\begin{align}
&\tilde{U}^\Pi \tilde{W}^\Pi \tilde{V}^{\Pi\dagger} \vec{y} = z 
\tilde{U}^\Theta \tilde{W}^\Theta \tilde{V}^{\Theta\dagger} \vec{y}.
\label{geig3}
\end{align}
Now we set 
\begin{align}
\vec{y} = \sum_{j=1}^{d} \xi_j \vec{v}^{\,\Pi}_j 
= \tilde{V}^{\,\Pi} \vec{\xi},
\end{align}
where $\tilde{V}^{\,\Pi} = (\vec{v}^{\,\Pi}_1,\cdots,
\vec{v}^{\,\Pi}_{d})$ and, then, 
the Eq. (\ref{geig3}) is rewritten as 
\begin{align}
\tilde{\Pi}\,\vec{\xi} &= z \,\tilde{\Theta}\,\vec{\xi},
\label{geig4}
\end{align}
where $\tilde{\Pi} \equiv \tilde{W}^{\,\Pi}$ and 
$\tilde{\Theta}\equiv \tilde{U}^{\,\Pi\dagger} \tilde{U}^\Theta 
\tilde{W}^\Theta \tilde{V}^{\Theta\dagger}
\tilde{V}^{\,\Pi}$.
Here $\tilde{\Pi}$ and $\tilde{\Theta}$ are square matrices of order $d$. 

Now the problem is reduced to the ordinary \lq\lq generalized\rq\rq\,eigenvalue problem and 
we can find the eigenvalues by solving 
${\rm det}\,(\,\tilde{\Pi}-z \,\tilde{\Theta})=0$ with respect to $z$. 
Suppose that $\vec{\xi}$ is the eigenvector satisfying Eq. (\ref{geig4}), 
that of the original problem, Eq. (\ref{geig}), is 
obtained from $\vec{x} = \tilde{V}\tilde{W}^{-1}\tilde{V}^{\,\Pi}\vec{\xi}$. 

One should note that zero or infinite eigenvalues may obtained in solving Eq. (\ref{geig4}). 
These eigenvalues originates from the remnant null space of $\tilde{\Pi}$ or $\tilde{\Theta}$. 
Actually, from Eq. (\ref{geig4}), the formula ${\rm det}\,(\,\tilde{\Pi}-z \,\tilde{\Theta})$
is a polynomial of degree $d$. 
If ${\rm rank}(\tilde{\Pi})<d$, a null vector exists for $\tilde{\Pi}$, which 
is an eigenvector belonging to eigenvalue zero. 
In this case, the zero-th order term of ${\rm det}\,(\,\tilde{\Pi}-z \,\tilde{\Theta})$ vanishes, 
resulting in a zero eigenvalue. 
If ${\rm rank}(\tilde{\Theta})<d$, a null vector exists for 
$\tilde \Theta$ and the $d$-th order term of ${\rm det}\,(\,\tilde{\Pi}-z \,\tilde{\Theta})$ vanishes. 
In this case, we have a eigenvalue of infinity\cite{infinit}, 
or in other words, ${\rm det}\,(\zeta\,\tilde{\Pi}-\tilde{\Theta})=0$ has the solution $\zeta=0$. 

In this section, we have limited the solution space of $\vec{x}$ in the Eq. (\ref{geig}), 
by deleting some vectors from the basis of $\vec{x}$. 
These vectors are not relevant for the transport phenomena, 
since they correspond to quantum states localized within a block and isolated from the 
neighboring blocks. 
The zero and infinite eigenvalues are also corresponding to 
the similar states and are safely neglected.

\section{Degeneracy in the Transfer Eigenvalues}
\label{degeneracy}

We have to pay attention to the cases, 
where the transfer eigenvalues $z$ are degenerate. 
In this case we have freedom of mixing the eigenvectors of 
the degenerate modes and then the Eq. (\ref{current-diagonal}) 
is not satisfied. 
Here we show how to deal with this situation. 

Let us denote a matrix composed by the degenerate wave functions by 
${\Phi}_D \equiv (\vec{\phi}_1^D,\cdots, \vec{\phi}_N^D)$, 
where $N$ is the degree of the degeneracy. 
We consider a matrix $\hat J$ given by 
$\hat{J} =(i/\hbar)\Phi_D^\dagger \Lambda \Phi_D$, 
which is diagonal if the Eq. (\ref{current-diagonal}) is hold true. 
Noting that $\hat{J}$ is symmetric, we can diagonalize $\hat{J}$ 
as $U^\dagger \hat{J} U = (i/\hbar)U^\dagger \Phi_D^\dagger \eta \Phi_D U$ 
by a unitary matrix $U$. 
Reselecting the eigenvectors in the degenerate subspace as 
$\Phi_D U = (\tilde{\vec{\phi}}_1^D,\cdots , \tilde{\vec{\phi}}_N^D)$,
we can make the matrix $\hat J$ diagonal. 

For evanescent modes the degeneracies are not harmful, 
since we do not utilize the property shown by the Eq. (\ref{current-diagonal})  
with respect to the evanescent modes in evaluating the Landauer formula. 


\begin{thebibliography}{10}
\bibitem{truncation}
We can take $d = {\rm max}({\rm rank}(\Pi), {\rm rank}(\Theta))$ instead. 
However the final results are essentially unchanged.
\bibitem{infinit}
For example, the solutions of the equation, $a x^2 + b x +c=0$, is $x=(-b\pm
  \sqrt{b^2-4 a c})/(2a)$. In the limit of $a\rightarrow 0$ the solutions tend
  to $x=-c/b$ and $x=-b/a$. Then the latter one diverges if $b\ne 0$.
\vfill\eject
\end{thebibliography}

\end{document}